\documentclass[a4paper,11pt]{article}
\usepackage{graphicx}
\usepackage[space]{grffile}
\usepackage{latexsym}
\usepackage{textcomp}
\usepackage{longtable}
\usepackage{tabulary}
\usepackage{booktabs,array,multirow,multicol,diagbox}
\usepackage[normalem]{ulem}
\usepackage{colortbl}
\usepackage{amsfonts,amsmath,amssymb}
\usepackage{natbib}
\usepackage{url}
\usepackage{hyperref}
\usepackage{etoolbox}
\makeatletter
\usepackage{authblk}
\usepackage[utf8]{inputenc}
\usepackage[english]{babel}
\usepackage[top=0.8in, bottom=0.8in, left=0.8in, right=0.8in]{geometry}

\title{Changes in Seasonal Upper Tropical Momentum Fluxes with Global Warming}

\author[1,2]{Abu Bakar Siddiqui Thakur  \thanks{Corresponding author: A.B.S. Thakur, thakur.abubakar@gmail.com}}
\author[1,2]{Jai Sukhatme}

\affil[1]{Centre for Atmospheric and Oceanic Sciences, Indian Institute of Science, Bangalore 560012, India}
\affil[2]{Divecha Centre for Climate Change, Indian Institute of Science, Bangalore 560012, India}

\begin{document}
	
\maketitle
% \clearpage
\begin{abstract}
    \normalsize{The boreal summer tropical upper-tropospheric momentum budget involves a balance between eddy and mean meridional fluxes. In winter, however, the eddy flux itself acts to accelerate and decelerate the zonal flow in the Asian and East Pacific regions, respectively. In a zonal mean sense, the residual of these two is then balanced by the mean meridional flux. These features are qualitatively captured by the CMIP6 suite of models in their control runs. With warming, the CMIP6 ensemble shows that the flux budget changes in a quantitative manner, in both the summer and winter seasons. Apart from the mean meridional flux which is affected by the projected weakening of the Hadley Cells, there are significant changes in eddy fluxes too. Notably, stationary wave fluxes are affected in the Asian and East Pacific regions during the summer and winter seasons, respectively. In the wintertime, extratropical wave activity penetrating into the East Pacific almost shuts off due to a weakening of the prevalent westerlies in the warming simulations. Whereas, in summer, eddy fluxes in the Asian region are displaced upward and changes are observed in the upper troposphere and lower stratosphere. Specifically, rotational flow around the Asian summer anticyclone weakens in the upper troposphere while strengthening in the lower stratosphere. Concomitantly, eddy flux convergence over the equatorial Indian Ocean decreases in the former and increases (by a larger amount) in the latter. In fact, strengthening of summertime tropical and subtropical stationary waves in the lower stratosphere with warming is not restricted to the Asian sector but is observed over all longitudes. Effectively, the magnitude of all terms in the upper tropospheric momentum flux budget decreases with warming, and an ensemble mean continues to yield a marginally westward annual and zonal mean equatorial zonal flow.} 
\end{abstract}

\section{Introduction}

The zonally averaged zonal momentum budget of the tropical upper troposphere has been the subject of numerous investigations \citep[see, for example,][]{lee1999climatological, dima2005tropical, kraucunas2005equatorial}.
The primary balance, as seen in reanalayis data, is between acceleration of the zonal mean flow by eddies and deceleration provided by the seasonally reversing Hadley circulation. 
In an annual mean, the easterly acceleration generated by the mean meridional advection of zonal momentum exceeds the eddy momentum flux convergence and leads to tropical upper tropospheric easterlies \citep{lee1999climatological}. 
Even on seasonal timescales, there is a strong sense of anti-correlation between these two terms \citep{dima2005tropical,kelly2011zonal}.
In fact, while emphasising on the roles of planetary and sub-planetary scale eddies in the winter-summer transition of the Northern Hemisphere (NH) circulation, a similar balance between mean meridional convergence and planetary-scale eddy momentum flux convergence was noted \citep{shaw2014role}.

This tendency for opposition between the two convergence terms occurs because of the preference of the zonal mean tropical rain belts and eddy forcing to occur in the same latitudinal band. 
The zonally averaged heating generates the meridional overturning circulation and a resultant upper tropospheric mass flux divergence, while the eddy forcing results in a momentum flux convergence into the source region. Apart from reanalysis, this line of reasoning is also supported by idealised modelling efforts \citep{kraucunas2005equatorial, kraucunas2007tropical}. The eddy momentum convergence is generated by climatological stationary Rossby gyres \citep{dima2005tropical, grise2011planetary, zurita2019role}, that are forced by longitudinal thermal contrasts \citep{wang1999seasonal, held2002northern}.
In fact, zonal asymmetries are prominent in the tropics with oceans and land masses, monsoon regions and deserts found in the same latitude band. In addition to the zonal asymmetry of the stationary waves, there is a growing body of research which acknowledges the zonally heterogeneous nature of the meridional circulation itself \citep{sun2019regional, raiter2020tropical}. This longitudinal structure, and potential variation of fluxes is lost when a zonally averaged picture of the tropical momentum budget is considered.

An understanding of this balance becomes all the more relevant in the context of anthropogenic climate change.
This is because, warming scenarios that have focused on understanding changes in the hydrological cycle have revealed a slow-down of the tropical circulation \citep{held2006robust, collins2013long, vallis2015response,ma2018responses}. 
Backed by observations \citep{seidel2008widening, nguyen2013hadley}, idealized modelling studies show that a poleward expansion of the tropical belt is a strong eventuality \citep{frierson2007width, levine2011response,  levine2015baroclinic}. 
Such changes in the large-scale Hadley circulation are likely to influence the mean flow deceleration of the zonal wind in the deep tropical upper troposphere. 
Further, many climate models develop an El Ni{\~n}o-like sea surface temperature (SST) anomaly in the equatorial Pacific in response to global warming \citep{dinezio2009climate, xie2010global}. 
As the thermal forcing is spread over a larger meridional extent, the atmospheric response is distinct from that to a normal El Ni{\~n}o event \citep{LuetalElNioversusGlobalWarming, tandon2013understanding}, and the overall effect is to zonalize the flow by weakening upper tropospheric stationary waves \citep{joseph2004global, LB1}. Further, intraseasonal activity of the Madden-Julian Oscillation increases in a warming world, and the associated eddy fluxes have been shown to push the zonal mean zonal flow towards a state of equatorial superrotation \citep{caballero2010spontaneous,arnold2012abrupt,carlson2016enhanced}
On another front, climate proxy data suggest that a similar El Ni{\~n}o-like SST pattern may have existed in past warm climates \citep{wara2005permanent, fedorov2006pliocene}.
This led to hypotheses regarding the prevalence of a permanent El Ni{\~n}o-like state maintained by equatorial superrotation, in the past and possibly in future warm periods \citep{pierrehumbert2000climate, tziperman2009pliocene}. Thus, projections of weakened Hadley Cell as well as geographically local overturning monsoonal circulations in warming scenarios \citep{ueda2006impact, sooraj2015global, wang2020understanding}, changes in stationary waves \citep{willis} and transient intraseasonal eddy activity may have important consequences for the momentum flux balance in the tropical atmosphere. 

Here, against this backdrop, we explore whether there are changes in the upper tropical momentum flux budget in a warming scenario. We start by highlighting that the present-day balance in the tropics is actually a little more delicate than what has been emphasized in literature. Specifically, we use reanalysis data and showcase the longitudinal structure of momentum fluxes as a function of time through the year (Section 3). We then move to CMIP6 control runs and projections under global warming. Section 4 begins with a comparison of CMIP6 control runs with reanalysis. We then discuss statistically significant changes in the equatorial mean meridional and eddy fluxes during winter and summer by comparing the control runs to a warming scenario. Changes in each season, their geographical origins and possible causes are then brought forth. In particular, it is seen that, in the equatorial region, during summer, stationary eddy fluxes are displaced upward and hence the influence of warming is felt in both upper tropospheric, and lower stratospheric momentum fluxes. The wintertime eddy flux changes are also significant though restricted to the upper troposphere.
Section 5 contains a summary of the results and their discussion.

\section{Data \& Methods}

The data used in the first part of this study comprises of daily-averaged horizontal winds at a resolution of $2.5^{\circ}$ across 17 pressure levels from ERA-Interim \citep{dee2011era} for a 40-year period from 1979-2018. We also make use of monthly mean GPCP Precipitation data \citep{adler2003version} with the same resolution and over the same period provided by NOAA (\url{https://psl.noaa.gov/}).

The zonally averaged zonal momentum equation reads \citep{dima2005tropical,kraucunas2005equatorial},
\begin{equation}\label{eq:1}
\frac{\partial [u]}{\partial t} = 
[v] \left(f - \frac{1}{\cos\phi}\frac{\partial [u] \cos\phi} {\partial y} \right) - \frac{1}{\cos^2 \phi}\frac{\partial [u^*v^*]\cos^2 \phi}{\partial y} - [\omega]\frac{\partial [u]}{\partial p} - \frac{\partial [u^* \omega^*]}{\partial p} + [\overline{X}]
\end{equation}

The notation above is standard. 
Specifically, square braces denote a zonal mean and asterisks denote a deviation from this mean. 
The first term on the right is the mean meridional momentum flux convergence, the second term is the meridional eddy momentum flux convergence. 
The third and fourth terms are the mean vertical advection and vertical eddy flux convergence. 
The last term is a residual and accounts for all sub grid-scale processes.
Eddies are computed as differences from the global mean.
For Day of Year variations in the zonal momentum budget, daily estimates of each term are calculated and then these are averaged over the respective days through the 40 years on record. 

For part of our analysis, we make use of a rotational-divergent partition.
Specifically, treating the horizontal wind field on each pressure level as a two-dimensional vector field, we split it into rotational and divergent components via a Helmholtz decomposition. 
The horizontal velocity field is expressed as, 
\begin{equation}\label{eq:2}
\Vec{v} = \nabla \chi - \Vec{k} \times \nabla \psi.
\end{equation}
The first term on the RHS describes the irrotational component while the second term on the RHS describes the non-divergent component of the velocity field. 
The rotational and divergent components will be denoted by the subscripts $r$ and $d$ respectively. 

\begin{table}[t]
	\centering
	\caption{
		List of models in the 1pctCO$_2$ and control simulation ensembles used in this study. The names of the variables are presented here as they appear in the IPCC nomenclature, and their availability is marked with a $\bullet$. Models highlighted with boldface are those for which daily horizontal wind data was available. \\}
    \resizebox{\columnwidth}{!}{%
    \begin{tabular}{|c|ccccc|c|ccccc|} 
    \hline
    \diagbox{\textbf{Models}}{\textbf{Data}} & \multicolumn{1}{c|}{\textbf{ua }} & \multicolumn{1}{c|}{\textbf{va}} & \multicolumn{1}{c|}{\textbf{zg }} & \multicolumn{1}{c|}{\textbf{ta}} & \textbf{tos }          & \diagbox{\textbf{Models}}{\textbf{Data}} & \multicolumn{1}{c|}{\textbf{ua }} & \multicolumn{1}{c|}{\textbf{va}} & \multicolumn{1}{c|}{\textbf{zg }} & \multicolumn{1}{c|}{\textbf{ta}} & \textbf{tos }           \\ 
    \hline
    \textbf{ACCESS-CM2}                      & $\bullet$            & $\bullet$           & $\bullet$            & $\bullet$           & $\bullet$ & GISS-E2-1-H                              & $\bullet$            & $\bullet$           & $\bullet$            & $\bullet$           & $\bullet$  \\
    \textbf{ACCESS-ESM1-5}                   & $\bullet$            & $\bullet$           & $\bullet$            & $\bullet$           & $\bullet$ & GISS-E2-2-G                              & $\bullet$            & $\bullet$           & $\bullet$            & $\bullet$           & $\bullet$  \\
    BCC-CSM2-MR                              & $\bullet$            & $\bullet$           & $\bullet$            & $\bullet$           & $\bullet$ & \textbf{IITM-ESM}                        & $\bullet$            & $\bullet$           & $\bullet$            & $\bullet$           & $\bullet$  \\
    BCC-ESM1                                 & $\bullet$            & $\bullet$           & $\bullet$            & $\bullet$           & $\bullet$ & MCM-UA-1-0                               & $\bullet$            & $\bullet$           & $\bullet$            & $\bullet$           & $\bullet$  \\
    \textbf{CanESM5}                         & $\bullet$            & $\bullet$           & $\bullet$            & $\bullet$           & $\bullet$ & \textbf{MIROC6}                          & $\bullet$            & $\bullet$           & $\bullet$            & $\bullet$           & $\bullet$  \\
    \textbf{CESM2}                           & $\bullet$            & $\bullet$           & $\bullet$            & $\bullet$           & $\bullet$ & \textbf{MPI-ESM-1-2-HAM}                 & $\bullet$            & $\bullet$           & $\bullet$            & $\bullet$           & $\bullet$  \\
    CESM2-FV2                                & $\bullet$            & $\bullet$           & $\bullet$            & $\bullet$           & $\bullet$ & \textbf{MPI-ESM-1-2-HR}                  & $\bullet$            & $\bullet$           & $\bullet$            & $\bullet$           & $\bullet$  \\
    \textbf{CESM2-WACCM}                     & $\bullet$            & $\bullet$           & $\bullet$            & $\bullet$           & $\bullet$ & \textbf{MPI-ESM-1-2-LR}                  & $\bullet$            & $\bullet$           & $\bullet$            & $\bullet$           & $\bullet$  \\
    \textbf{CESM2-WACCM-FV2}                 & $\bullet$            & $\bullet$           & $\bullet$            & $\bullet$           & $\bullet$ & \textbf{MRI-ESM2-0}                      & $\bullet$            & $\bullet$           & $\bullet$            & $\bullet$           & $\bullet$  \\
    \textbf{CMCC-CM2-SR52}                   & $\bullet$            & $\bullet$           & $\bullet$            & $\bullet$           & $\bullet$ & NorCPM1                                  & $\bullet$            & $\bullet$           & $\bullet$            & $\bullet$           & $\bullet$  \\
    \textbf{CMCC-ESM2}                       & $\bullet$            & $\bullet$           & $\bullet$            & $\bullet$           & $\bullet$ & \textbf{NorESM2-LM}                      & $\bullet$            & $\bullet$           & $\bullet$            & $\bullet$           & $\bullet$  \\
    FGOALS-g3                                & $\bullet$            & $\bullet$           & $\bullet$            & $\bullet$           &                        & \textbf{NorESM2-MM}                      & $\bullet$            & $\bullet$           & $\bullet$            & $\bullet$           & $\bullet$  \\
    FIO-ESM-2-0                              & $\bullet$            & $\bullet$           & $\bullet$            & $\bullet$           & $\bullet$ & SAM0-UNICON                              & $\bullet$            & $\bullet$           & $\bullet$            & $\bullet$           & $\bullet$  \\
    GISS-E2-1-G                              & $\bullet$            & $\bullet$           & $\bullet$            & $\bullet$           & $\bullet$ & TaiESM1                                  & $\bullet$            & $\bullet$           & $\bullet$            & $\bullet$           &                         \\
    \hline
    \end{tabular}
    }
    \label{data_tab}
\end{table}

In addition, for estimates of the future climate, we make use of full complexity model simulations from the CMIP6 archive \citep[\url{https://esgf-node.llnl.gov/projects/cmip6/}]{eyring2016overview}.
In particular, we employ 28 model simulations forced using the $1\%$yr$^{-1}$ increase in CO$_2$ concentration (\textit{1pctCO2}), along with the pre-industrial control simulation (\textit{piControl}), for the assessment of the impact of anthropogenic climate change on tropical momentum fluxes.
For each model we use the first (\textit{r1i1p1f1}) ensemble member.
The models and data used are listed in Table \ref{data_tab}.
Prior to calculations, data from each model is interpolated onto a common $2.5^{\circ}\times2.5^{\circ}$ grid.
The calculations presented here are performed over years 131 through 150 for both the forced and control experiments.
Flux calculations are performed using daily data.
Other results presented use computations with monthly data or by down sampling from daily data.
Statistical significance of the changes is tested using a two-tailed Student's t test for difference between two independent sample means, with the assumption of unequal variances.

\section{Present-day upper tropical momentum fluxes}

\begin{figure}[t]
	\centering
	\includegraphics[width=0.9\linewidth]{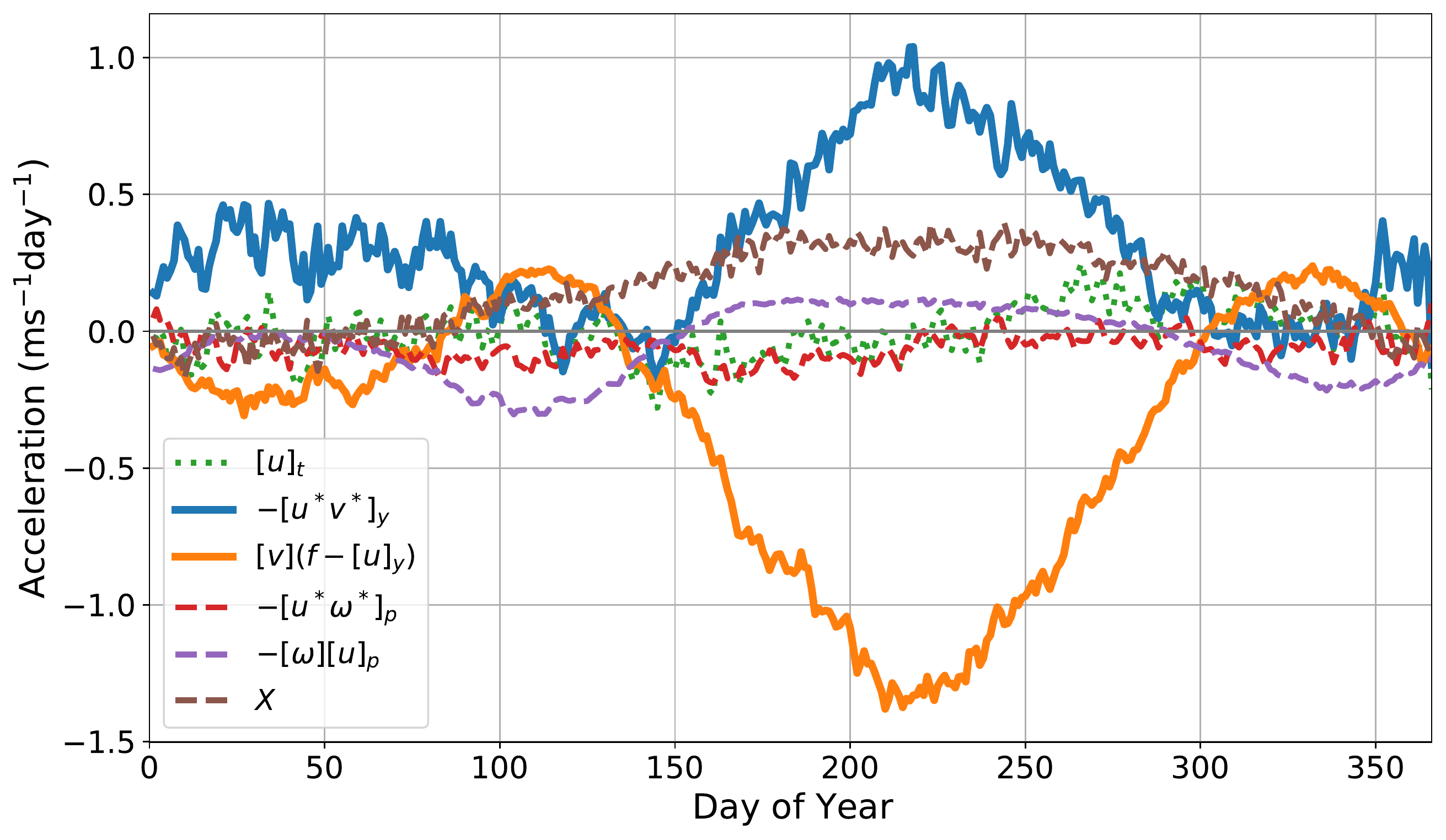}
	\caption{Climatological Day of Year variation of each term of the zonally averaged zonal momentum budget, Equation \protect \ref{eq:1}, averaged over 150-300 hPa, $\pm 5^{\circ}$ of the equator.}
	\label{fig:era_mom}
\end{figure}

The annual cycle of the zonal mean zonal momentum equation, i.e., Equation \ref{eq:1} for the tropical upper troposphere is presented in Figure \ref{fig:era_mom}. 
Consistent with previous studies that focused on yearly or seasonal means, the eddy and mean meridional momentum convergences lead the momentum budget of the tropical upper troposphere. Compared to the rest of the year, both these terms are particularly enhanced during the Asian summer monsoon season (June through September) and are much smaller during the equinoctial seasons.
During summer, the residual term ($\overline{X}$; brown dashed line) is relatively larger in magnitude ($\sim$0.4 ms$^{-1}$day$^{-1}$) than the other remaining terms, due to strong convective momentum transport over Indian Ocean - Maritime Continent region \citep{lin2008sources}.
Thus, the discussion in the rest of this paper will focus on the mean and eddy momentum flux convergence terms.

\begin{figure}[t]
	\centering
	\includegraphics[width=0.9\linewidth]{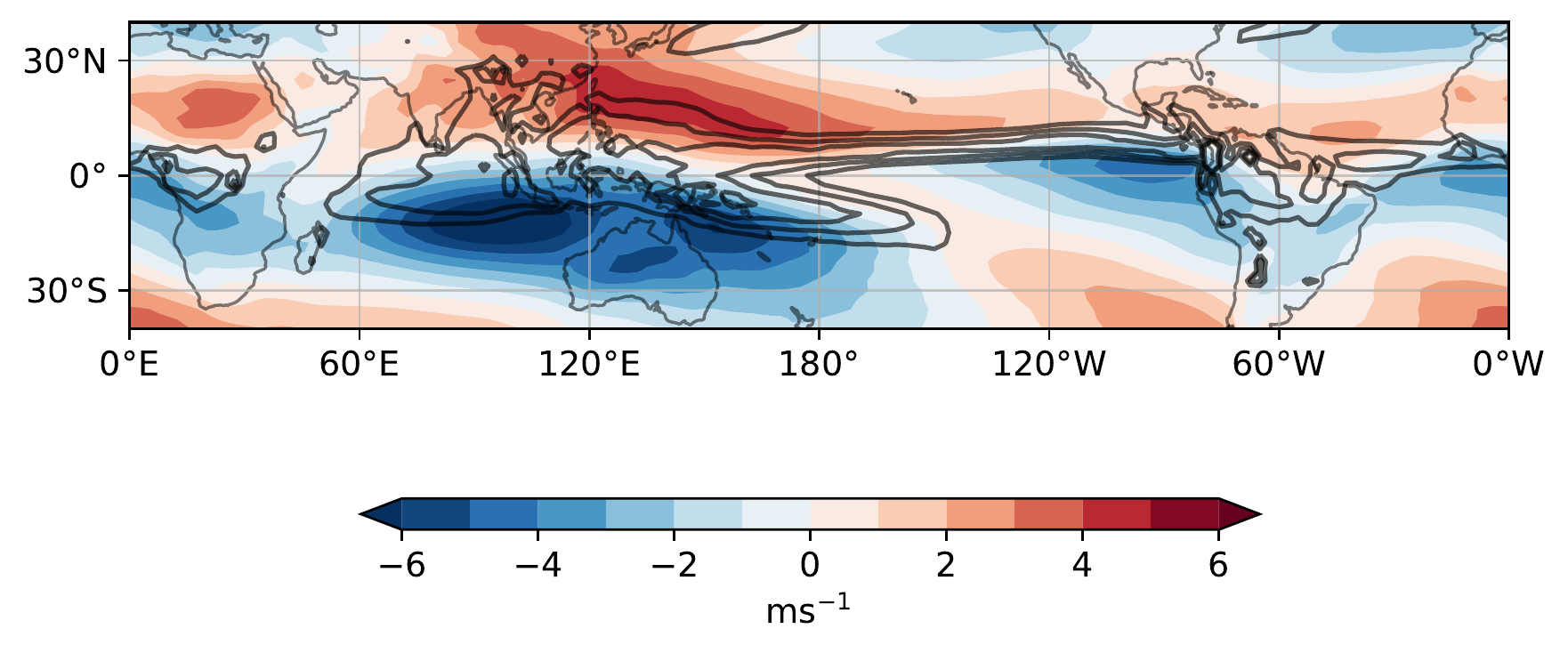}
	\caption{Spatial map of the annual mean difference between upper (150 hPa) and lower level (925 hPa) divergent meridional winds (colors), along with time mean precipitation (contours). The contours of precipitation are at 5, 7, 9 and 11 mm/day.	
	}
	\label{fig:diffVd}
\end{figure}

Recent work highlights that the seasonally varying zonally averaged mean meridional circulation is actually composed of longitudinally limited overturning circulations \citep{hoskins2019detailed}.
Following the argument that it is the divergent motions that contribute to the north-south meridional overturning circulations \citep{zhang2013interannual, schwendike2014local}, we construct a spatial map of the difference between the upper and lower tropospheric divergent meridional wind as an annual mean overlaid with contours of annual mean precipitation (Figure \ref{fig:diffVd}).
Such a difference emphasizes the upper level divergent motion forced by the monsoonal heating.
Based on Figure \ref{fig:diffVd}, the tropical overturning activity can be split largely, into two regions. These are the Asia-Africa region (30W - 150W; abbreviated as Af-A) and Central Pacific - West Atlantic (150W - 30W; abbreviated as CP-WA) region. 
These partitions are motivated by the localised monsoonal circulations that exist within them.

\begin{figure}[t]
	\centering
	\includegraphics[width=0.9\linewidth]{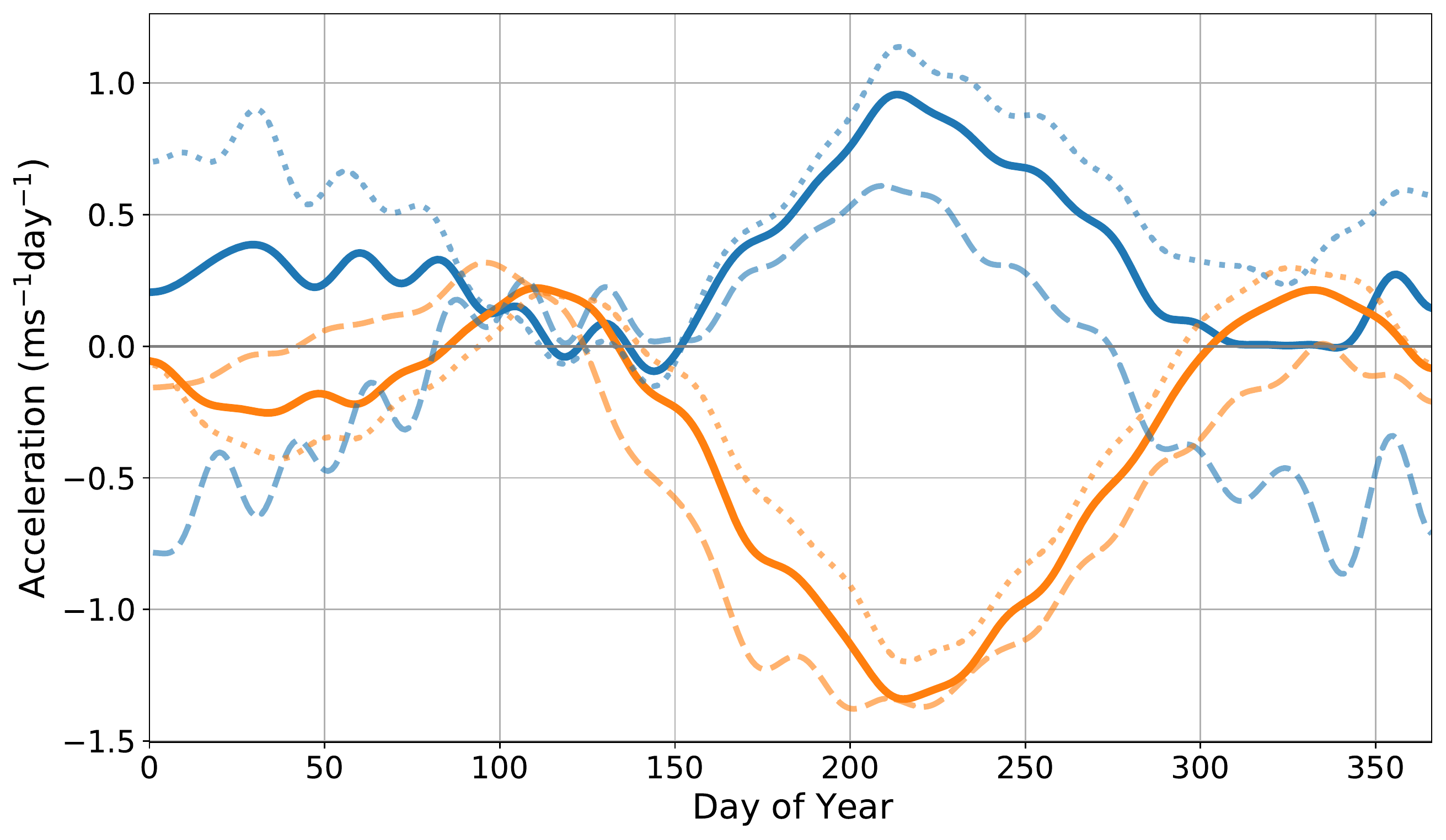}
	\caption{Same as Figure \protect \ref{fig:era_mom} except that the zonally averaged eddy momentum flux calculations are over respective longitudinal bands mentioned in reference to Figure \protect \ref{fig:diffVd}. The eddies are computed as a difference from the global mean. The solid lines are representative of the entire domain. The dotted and dashed lines are for the Asia-Africa (Af-A) and Central Pacific - West Atlantic (CP-WA) sectors, respectively. A 20-day low-pass filter is applied prior to presentation.}
	\label{fig:eddy2zones}
\end{figure}

\begin{figure}[t]
	\centering
	\includegraphics[width=0.9\linewidth]{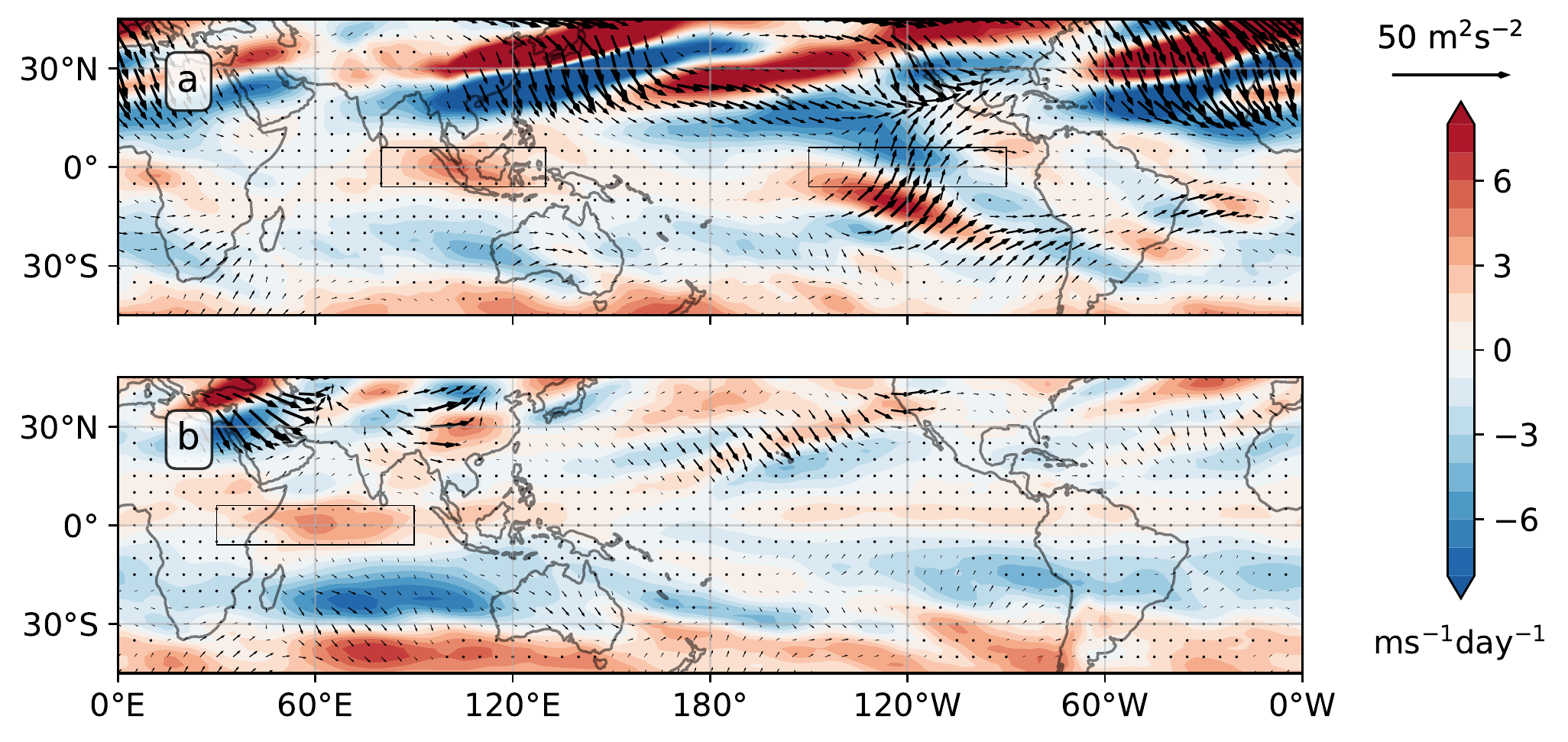}
	\caption{Upper tropospheric wave activity flux (arrows; at 250 hPa) along with eddy momentum flux convergence (colors; averaged over 150-300 hPa) for (a) winter (DJF) and (b) summer (JJA). The wave activity flux is computed following \cite{plumb1985three, caballero2009impact}. }
	\label{fig:eddiesClimWinSum}
\end{figure}

Considering these two longitudinal sectors, Figure \ref{fig:eddy2zones} show their contributions to each of the leading terms of Equation \ref{eq:1}, and Figure \ref{fig:eddiesClimWinSum} shows a spatial map of the momentum flux convergence.
Quite strikingly, while eddies in the Af-A sector tend to accelerate the zonal flow over the course of both the solstitial seasons, eddies in the CP-WA region actually decelerate the zonal flow during the winter season. 
When computed globally, this leads to a near cancellation between the two sectors during the winter (solid blue curve in Figure \ref{fig:eddy2zones}). The geographical regions responsible for this aforementioned cancellation are highlighted by black boxes in Figure \ref{fig:eddiesClimWinSum}a and Figure S1a. 
The positive eddy convergence in the 120E box is linked to the two off-equatorial anti-cyclonic Rossby gyres straddling the equator around the Maritime Continent (Figure S1a) of which the Southern Hemispheric (SH) gyre is tied to the Australian Monsoon.
These are the climatological stationary Rossby waves, forced by longitudinally dependent thermal contrasts and are present in this region throughout the year \citep{wang1999seasonal, dima2005tropical, kraucunas2005equatorial}. In contrast, the divergence of eddy momentum flux in the East Pacific is due to a convergence of extratropical wave activity generated in SH Pacific ocean \citep[120W box of Figure \ref{fig:eddiesClimWinSum}a; see also Figure 6 of][and the references therein]{barnes2012global}. Specifically, extra-tropical Rossby waves and the associated wave activity flux propagates across the equator into NH through the "westerly window" in and around the 120W box \citep{hoskins1993rossby, li2015interhemispheric}. 
In a zonal mean sense, the eddy acceleration is dominated by correlations between $u_r^*$ and $v_d^*$ \citep[see][and Figure S2a]{zurita2019role}. 
In winter, this zonal mean character is largely forced by Af-A region (Figure S2b) because the CP-WA area is dominated by extra-tropical rotational eddy fluxes \citep[$u_r^* v_r^*$, see][and Figure S2c]{zurita2019role}.

In the boreal summer, eddy fluxes in both the zones are accelerating in nature. For Af-A, the dominant feature responsible for the eddy flux convergence in the deep tropics is the upper tropospheric return flow of the Asian summer monsoon \citep[and Figure S1b]{dima2005tropical, hoskins2019detailed}, highlighted by the box in Figure \ref{fig:eddiesClimWinSum}b. 
In comparison to winter, the Pacific ocean gyres have a much weaker tilt over the East Pacific region during summer and the patch of westerlies becomes feeble (Figure S1b).
Consequently, there is no cross-equatorial propagation of extra-tropical Rossby waves or related eddy momentum flux divergence in the eastern tropical Pacific. 
Thus, rather than a deceleration, the flux convergence in this region is also positive during summer ($\sim$ 1-2 ms$^{-1}$day$^{-1}$) and is possibly related to the localized heat source due to the summer inter-tropical convergence zone \citep{suarezduffy,kraucunas2005equatorial}.

With regard to the mean meridional fluxes, a striking feature of Figure \ref{fig:eddy2zones} is the high degree of similarity (orange curves) of this term when computed over the entire globe and the two individual sectors.
The physical reason behind this feature is that the $u_r$-$v_d$ contribution and total mean convergence (solid and dashed orange curves of Figure S2) are identical. 
On splitting Equation \ref{eq:2} into its zonal and meridional wind components, it can be seen that that $v_r$ and $u_d$ are zonal gradients of a stream function and velocity potential, respectively, and go to zero when integrated over all longitudes. 
Further, as seen in Figures S2 and S3, the dominance of the $u_r$-$v_d$ contribution holds individually in both regions.
It may also be noted that $-[v_d]\partial_y[u_r]$ is simply the advection of zonal mean relative vorticity by the zonal mean divergent motions from the summer hemisphere into the winter hemisphere.
This theme of cross-equatorial transport of vorticity by the thermally direct circulation is consistent across all regions (Figure S3) and explains why there exists a strong similarity, in this regard, amongst the two sectors.

In all, the eddy momentum fluxes are quite diverse in nature. The Af-A sector provides the bulk of the eddy acceleration during both the solstitial seasons, while the CP-WA sector undergoes a semi-annual reversal. 
During winter, fluxes in the Af-A (CP-WA) are of tropical (extra-tropical) origin and try to accelerate (decelerate) the zonal flow. Whereas in summer, both these regions experience acceleration due to the convergent tendency of the eddy fluxes during this season. These are balanced by the negatively signed mean meridional momentum flux convergence. The magnitude of the mean term in each zone peaks in the boreal summer, despite representing different longitudinal zones.
Further, the nature of this mean flux convergence is similar for both the individual regions and the zonal mean throughout the year. With these considerations, it becomes evident that the tropical momentum balance is quite delicate and involves a three-way compensation between tropical eddy acceleration, extra-tropical eddy deceleration, and zonal-mean absolute vorticity advection by the divergent meridional flow.
A change in any of these three components has the potential to alter the upper tropical momentum budget. 
Given the many changes anticipated in the tropical circulation due to global warming, in the next section we explore if there is any discernible influence on these fluxes, and whether there is a change in the present-day tropical momentum balance in a warming scenario.

\section{Momentum fluxes in a warming scenario}

To begin with, Figure S4a shows the upper tropospheric eddy and mean momentum flux convergences for the control simulation of CMIP6.  
When compared against present-day reanalysis (Figure \ref{fig:eddy2zones}), the control set is able to qualitatively capture the zonal mean fluxes as well as the mean meridional convergence terms in both Af-A and CP-WA zones. 
Further, winter-time accelerating and decelerating nature of eddy fluxes in Af-A and CP-WA regions, respectively, is also captured by the the suite of CMIP6 models. While of the correct sign, one of the main discrepancies seen is in the estimates of the eddy acceleration during summer. 
Specifically, the control run has a weaker zonal acceleration than that seen in the present-day reanalysis. Secondly, we also note that the mean meridional deceleration in the CP-WA sector is larger in the control run than in reanalysis. 
A reason for these discrepancies is that the model fluxes tend to be a little displaced meridionally and vertically than the latitudinal band and levels used to analyze the present-day reanalysis. 
Given the overall qualitative agreement between the control and present-day estimates, it is worth looking at the influence of climate change. 

\begin{figure}[ht]
	\centering
	\includegraphics[width=0.9\linewidth]{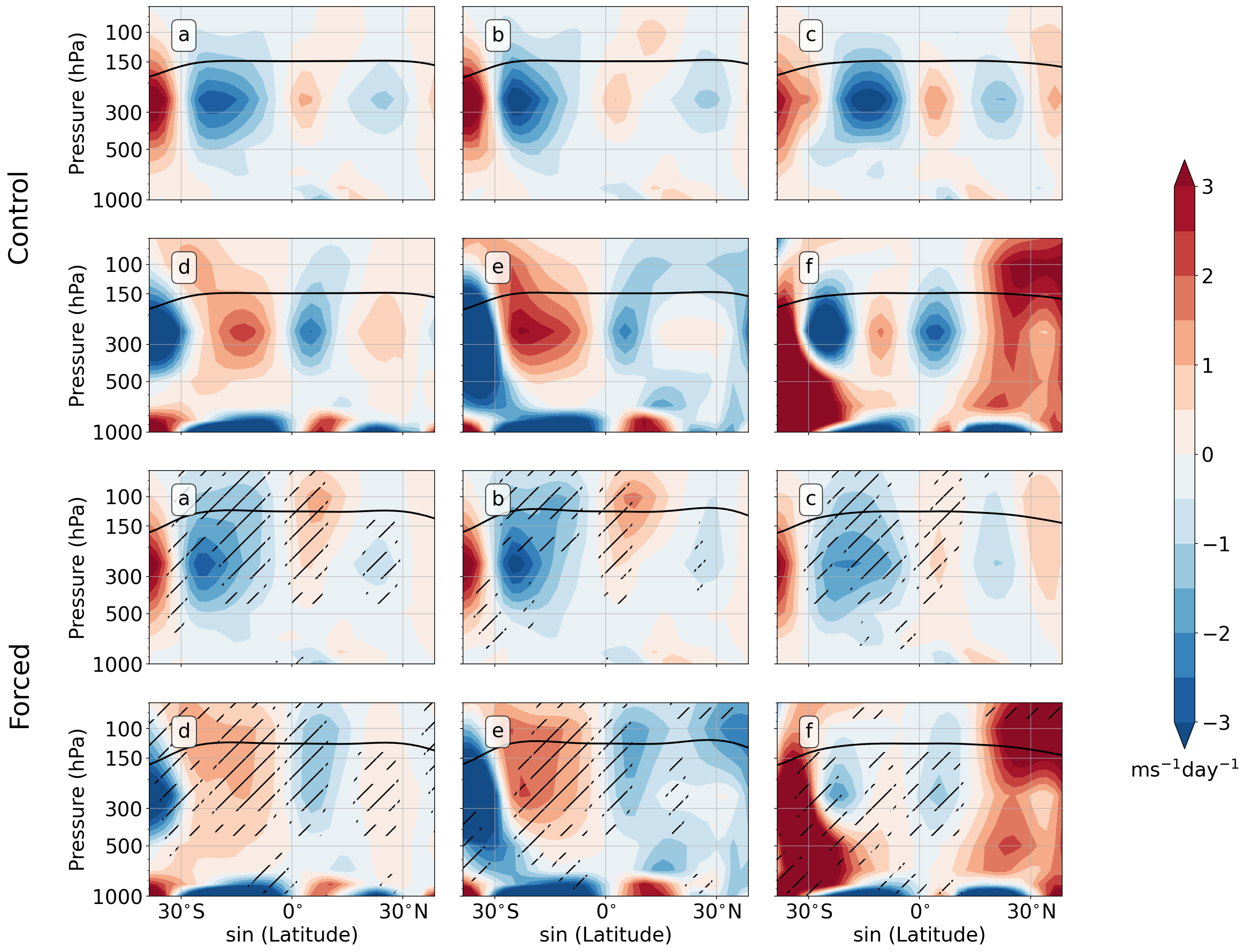}
	\caption{
		Seasonally and zonally averaged multi-model mean vertical structure of (a) eddy momentum flux convergence and (d) mean meridional momentum flux convergence for the CMIP6 (top half) control and (bottom half) forced simulations for summer (JJA).
		Also displayed alongside in each row are the respective quantities computed over (b,e) Af-A and (c,f) CP-WA. 
		In each panel, the black curve is indicative of the multi-model mean tropopause height over the corresponding region \protect \citep[see][]{reichler2003determining}.
		Hatching denotes changes that are statistically significant at the 95\% level by a two-tailed t-test.
	}
	\label{fig:cmip_mom_vert_sum}
\end{figure}

\begin{figure}[ht]
	\centering
	\includegraphics[width=0.9\linewidth]{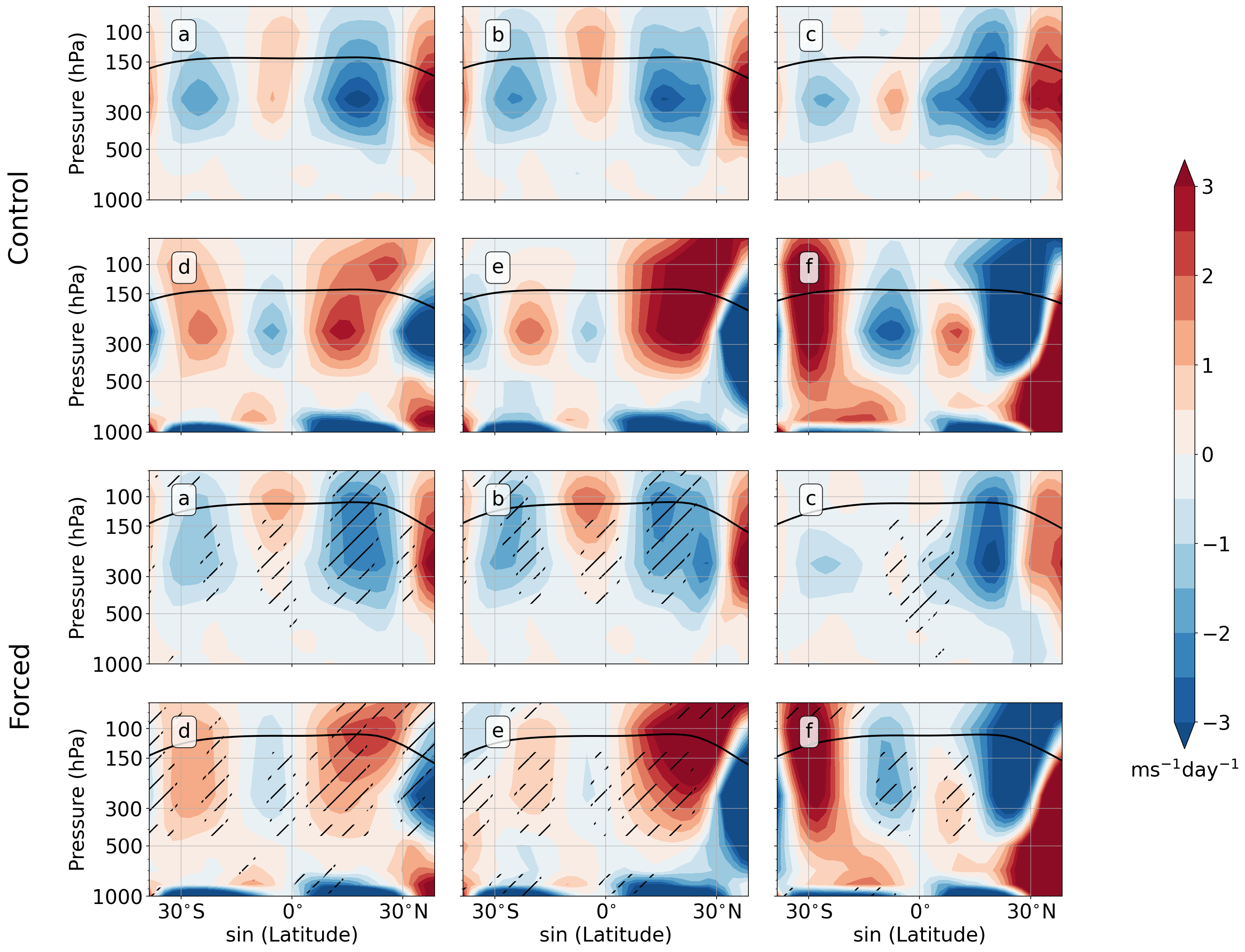}
	\caption{
		Same as Figure \protect \ref{fig:cmip_mom_vert_sum} except for seasonal mean over winter.
	}
	\label{fig:cmip_mom_vert_win}
\end{figure}

Indeed, fluxes of the the 1\% increase in CO$_2$ concentration runs with respect to the control run (Figure S4b) clearly suggest changes in both the summer and winter seasons.  
For example, the flux due the mean meridional term weakens considerably (though still decelerating in nature) over the CP-WA region. Even though these anomalies are noticeable primarily over the CP-WA sector, by extension, they are also manifest in globally averaged terms. Along with changes in the mean meridional term, during summer, we also notice there are changes in the eddy acceleration in the CP-WA and Af-A regions. In fact, overall, the magnitude of momentum fluxes decreases in the warming scenario in a zonal mean as well as in individual zonal sectors.
One of the most robust responses of anthropogenic climate change is an increase in the height of the tropopause \citep{lorenz, vallis2015response}.
As eddy fluxes of momentum tend to be concentrated in the upper troposphere \citep{ait2015eddy}, there arises a possibility wherein fluxes undergo an upward displacement along with the tropopause. Further, there is also scope for changes in the latitudinal profile of the fluxes.  
Thus, instead of layer and specific meridional band averaged line plots, such as those in Figures \ref{fig:eddy2zones} and S4, a view of the fluxes with height and latitude is more useful; this vertical-zonal mean section is shown in Figures \ref{fig:cmip_mom_vert_sum} and \ref{fig:cmip_mom_vert_win} for summer and winter seasons, over the global domain, Af-A and CP-WA regions, respectively.
As with the upper layer averages (Figure \ref{fig:eddy2zones}), the latitude-height computations for the control set bears resemblance to reanalysis \citep{dima2005tropical}.
However, the magnitudes of the terms are smaller in the control set than those of reanalysis.

Consistent with expectations of an increase in tropical tropopause height (marked by thick dark lines in Figures \ref{fig:cmip_mom_vert_sum} and \ref{fig:cmip_mom_vert_win}), the near-equatorial eddy acceleration shifts upward and gets concentrated near the 100 hPa level in the forced set (Figure \ref{fig:cmip_mom_vert_sum}a).
There is a clear and significant increase in the magnitude of the eddy flux in the Af-A sector in summer. Moreover, changes in the summer-time eddy flux in this region is observed in both the upper troposphere as well as the lower stratosphere (Figure \ref{fig:cmip_mom_vert_sum}b) . 
A similar increment is visible over this region in the winter season too (Figure \ref{fig:cmip_mom_vert_win}) --- though, here both the control and warming run fluxes are spread across the tropopause.
In comparison, there appears to be no vertical movement in the CP-WA sector during either season, i.e., the fluxes remain in the upper troposphere (Figures \ref{fig:cmip_mom_vert_sum}c and \ref{fig:cmip_mom_vert_win}c).
In fact, the magnitude of the eddy flux over this region decreases in the climate change forced scenario in both seasons. Considering the mean meridional deceleration (Figure \ref{fig:cmip_mom_vert_sum}d), its weakening is in line with that noted in Figure S4.
Comparatively weaker mean meridional fluxes can also be seen in the Af-A zone during summer (Figure \ref{fig:cmip_mom_vert_sum}e) and this decrease is most pronounced over CP-WA in the summer (Figure \ref{fig:cmip_mom_vert_sum}f).
Changes of a similar nature, though smaller in magnitude, are also noticeable in the winter season (Figure \ref{fig:cmip_mom_vert_win}). 
Interestingly, these weaker fluxes also appear to be spread over a deeper vertical layer in the forced runs. 

The decrease in strength of the mean meridional flux in the upper tropical region is anticipated by theories that posit a wider and weaker Hadley cell in a warming world \citep{lu2007expansion, levine2011response, vallis2015response}. 
On the other hand, the changing nature of eddy fluxes, their regional dependence and signature in the upper troposphere as well as lower stratosphere is intriguing and has not received much attention.
We now probe the cause of these changes in zonally averaged eddy fluxes on geographically localized scales.

\subsection{Modification of Eddy fluxes in the warming runs}

\begin{figure}[t]
	\centering
	\includegraphics[width=\linewidth]{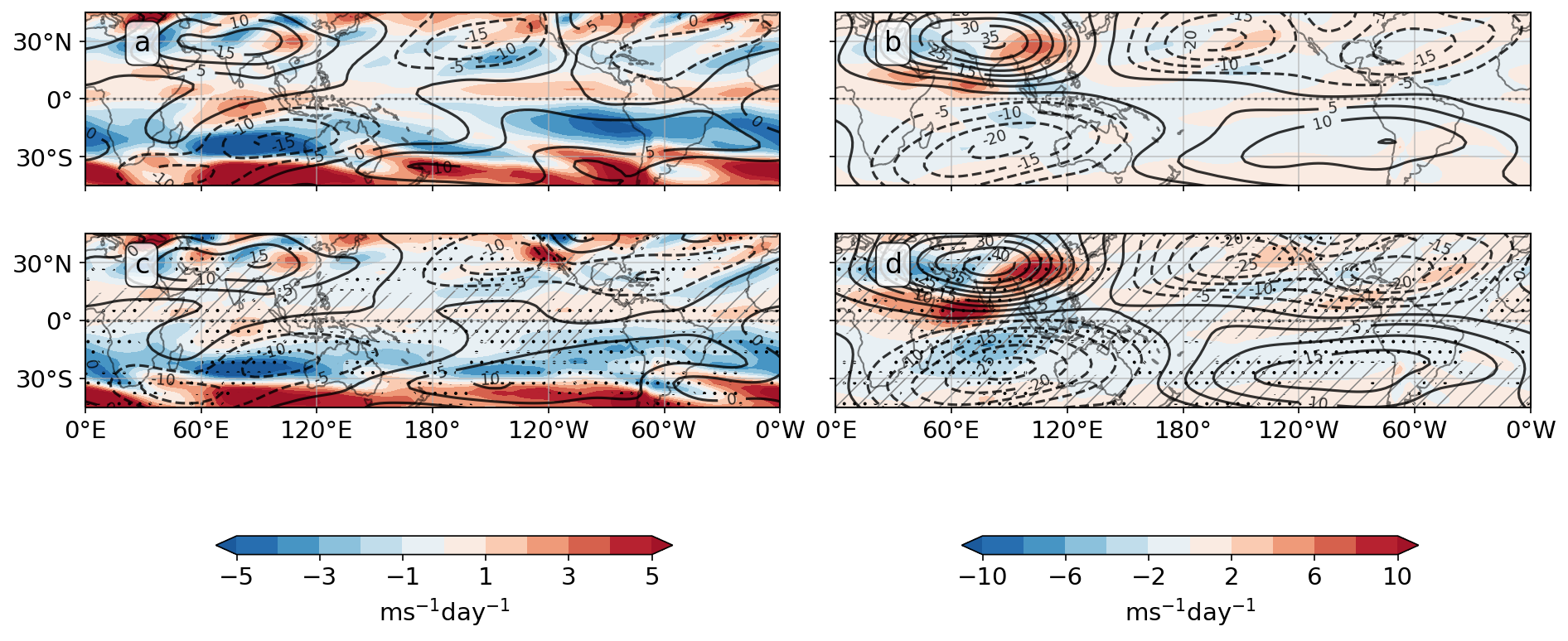}
	\caption{
		Boreal summer seasonal mean upper tropospheric (a,c; 250 hPa) and lower stratospheric (b,d; 100 hPa) spatial maps of eddy momentum flux convergence (colours) and eddy streamfunction (contours).
		Top (a,b) panels show the ensemble mean for control while the bottom (c,d) shows that for forced simulations.
		Contour intervals are the same in all panels and the units are $10^6$ m$^2$s$^{-1}$.
		Hatching with gray lines (stippling with black dots) denotes changes in streamfunction (eddy momentum flux convergence) that are statistically significant at the 95\% level by a two-tailed t-test
	}
	\label{fig:cmip_mey_sf}
\end{figure}

To probe the changes during summer in greater detail, we examine maps of the eddy momentum fluxes in the upper troposphere (250 hPa) and lower stratosphere (100 hPa) separately. As seen in Figures \ref{fig:cmip_mey_sf}a,c, the streamlines at 250 hPa indicate weakening and a more zonal nature of the subtropical summer stationary waves with warming \citep{joseph2004global,willis}. With regard to the momentum flux, the region which experiences the most change is Af-A, and here too we note a reduction in strength of the flow around the Asian monsoon anticyclone which is consistent with projections of a weaker monsoon circulation with global warming \citep{ueda2006impact, sooraj2015global, wang2020understanding}. Moreover, this weakening of the asymmetric circulation at upper tropospheric levels in warmer climates is attributed to the effect of smaller sea-surface temperature gradients that outweighs the increase in strength by uplift of the tropopause and associated increase in upward mass flux \citep{LB1}. In fact, as is seen via the colors in Figures \ref{fig:cmip_mey_sf}a,c, the momentum flux convergence actually decreases in the equatorial upper troposphere in this region. On the other hand, Figures \ref{fig:cmip_mey_sf}b,d suggest a strengthening of the subtropical stationary waves in the lower stratosphere (100 hPa). Specifically, at this level, flow around streamlines of the Asian monsoon anticyclone increases in strength with warming and so does the momentum flux convergence over the equatorial Indian Ocean. Taken together, the increase of eddy momentum flux in the lower stratosphere is larger than the decrease in the upper troposphere as is reflected in Figure S4 and Figure \ref{fig:cmip_mom_vert_sum}. 
Moreover, as seen in Figure \ref{fig:cmip_mey_sf}b,d, the summertime increase in strength of stationary waves in the lower stratosphere is evident across all longitudes in the subtropics and tropics. Clearly, stronger rotational flows are observed around these stationary waves in both the eastern and western hemispheres.

It has been noted that the wintertime stationary wave amplitudes and phases in the northern hemisphere upper stratosphere are expected to change with greenhouse gas forcing \citep{wang1}, and similarly extratropical southern hemisphere stationary waves have been noted to increase in magnitude in recent decades. The former appears to be linked to changing zonal mean flows in the stratosphere \citep{wang1}, while the latter is mainly due to an influence of ozone depleting substances \citep{wang2}. Changes in the lower stratospheric zonal mean flow during summer in the subtropics have been noted in warming simulations \citep{lorenz}, and further, the increase in magnitude of these zonally asymmetric anomalies that we observe appear to go hand in hand with the projected stronger zonal mean Brewer-Dobson circulation \citep{butchart,butchart1}. This connection could possibly be mediated via the contribution of the so-called "tropospheric control" of the lower stratospheric overturning circulation by stationary waves \citep{gerber}. Moreover, the increased mass flux from the troposphere to the stratosphere expected with warming during the northern hemisphere summer in the subtropics \citep{deckert}, plausibly from the Indian summer monsoon, is likely to result in a larger local lower stratospheric divergence and hence a stronger rotational wind via Sverdrup balance \citep{LB1}. In fact, comparing Sverdrup balance at 100 hPa (Figure S5) clearly indicates stronger divergence in the forced ensemble as compared to the control simulation. Similar behaviour is also observed (not shown) in simple aquaplanet models mimicking the planetary-scale boreal summer monsoon flow \citep{wu2016impact}, when subjected to a spatially uniform SST increase of 4K \citep{webb2017cloud}.

\begin{figure}[ht]
	\centering
	\includegraphics[width=\linewidth]{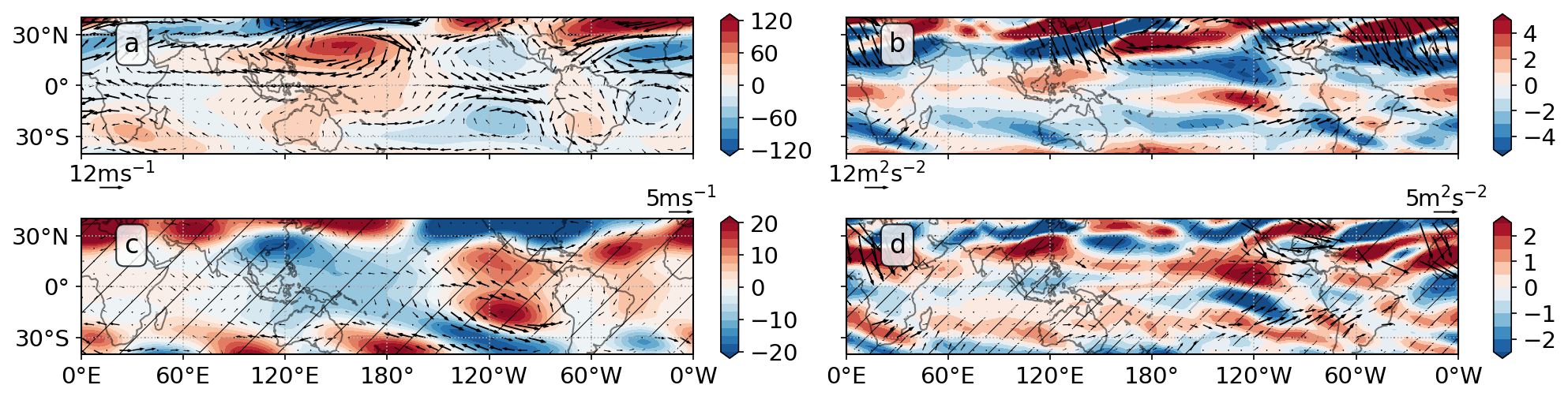}
	\caption{
		Boreal winter seasonal mean distribution of eddy geopotential height (m; left) and eddy momentum flux convergence (ms$^{-1}$day$^{-1}$; right), both for control (a,b) and difference between forced and control simulations (c,d), averaged at 250 hPa.
		Quivers represent eddy wind vectors (left) and wave activity flux vectors (right) computed at the same pressure level.
		In the climate change anomaly panels (c,d), the quivers are plotted at any point only if either $x$- or $y$- component of the respective quantity is statistically significant at 95\% level at that point.
		For the top (bottom) panels, the quiver keys are illustrated at the bottom-left (top-right) of the respective panel.
		Hatching denotes areas where the geopotential height (eddy momentum flux convergence) are statistically significant at the 85\% level by a two-tailed t-test.
	}
	\label{fig:cmip_zg_mey}
\end{figure}

With regard to the winter season, as noted, the most noticeable change occurs over the East Pacific region.
In fact, the decrease in magnitude of eddy flux in the CP-WA region (Figure \ref{fig:cmip_mom_vert_win}) is in the upper troposphere and mainly because of the weakening of the eddy deceleration features in the 120W box. Indeed, the cross-equatorial propagation of wave-activity, sizeable in reanalysis and subdued in the control set (Figure \ref{fig:cmip_zg_mey}b), is almost completely absent in the global warming projection (Figure \ref{fig:cmip_zg_mey}d). The change in stationary waves in this region is captured in Figure \ref{fig:cmip_zg_mey}c via  the appearance of a pair of equatorially symmetric anti-cyclonic gyres in the East Pacific.
The equatorial easterlies associated with these solitary highs interfere with the winter eddy westerlies prevalent in this region which, as noted before, are important for the cross-equatorial propagation of extra-tropical waves \citep{hoskins1993rossby}. 
% Not so apparent, yet statistically significant, are anomalies over the west coast of Africa.
% These negative anomalies are to the tune of $\sim$1-2 ms$^{-1}$day$^{-1}$ and occur over a small longitudinal stretch ($\sim$10$^\circ$ in extent), hence may not contribute significantly to the zonally averaged momentum budget. , 
Further, these wintertime gyres in the East Pacific are coincident with the ubiquitous El Ni{\~n}o-like SST anomaly simulated by climate models \citep[see Figure S6;][]{dinezio2009climate, xie2010global}.
In fact, atmosphere-only simulations forced by a single agent \citep{webb2017cloud} reveal that these changes in the upper tropical troposphere (250 hPa) can largely be attributed to the indirect impact of anthropogenic climate change via SST boundary forcing (Figure S7). Most of the features described for the upper troposphere during summer and winter are reproduced in the boundary forcing experiments. 

\begin{figure}[t]
	\centering
	\includegraphics[width=0.6\linewidth]{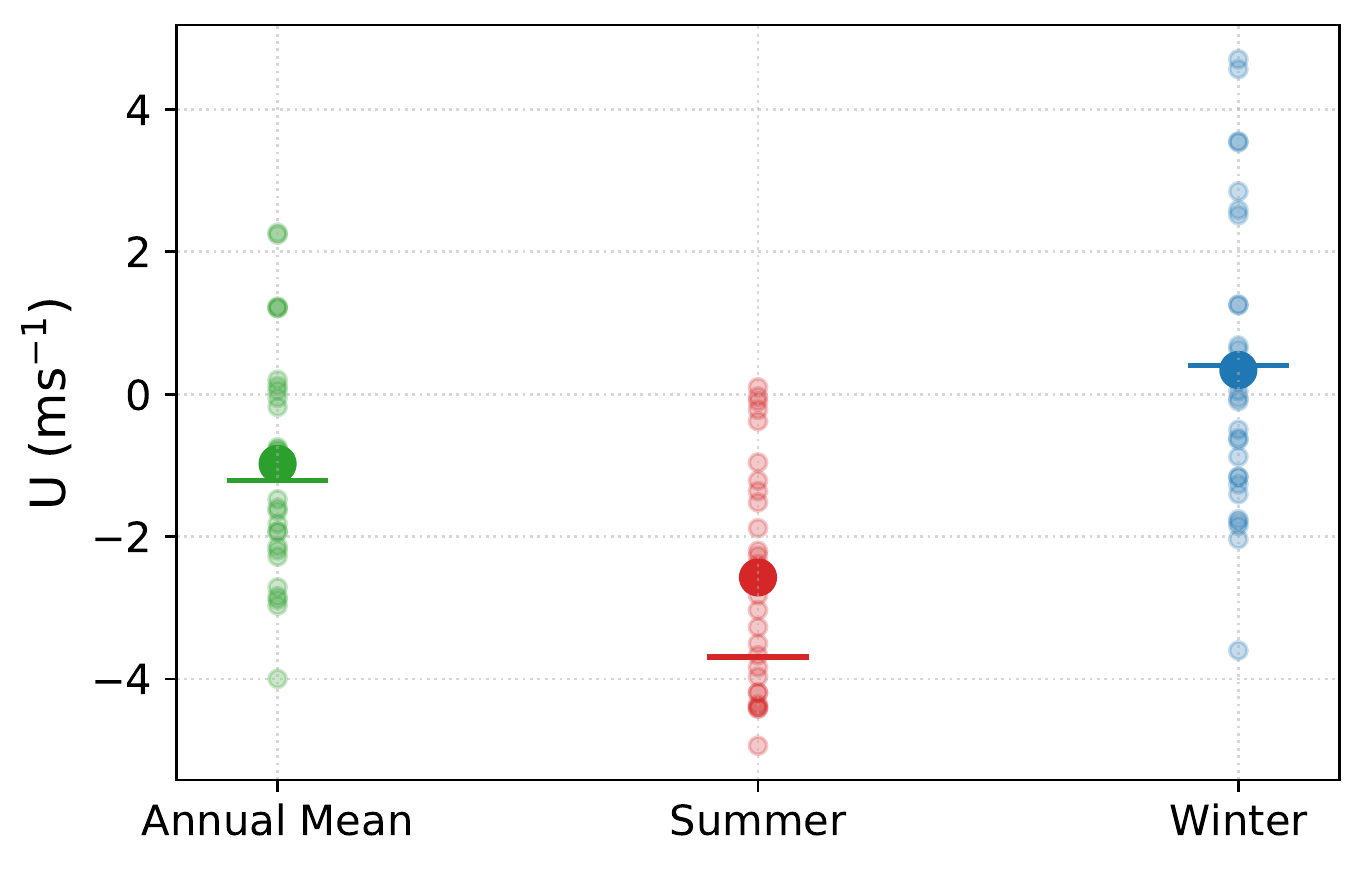}
	\caption{Time mean zonal mean zonal wind spread for the CMIP6 fully coupled simulations at 250 hPa, averaged over $\pm5^{\circ}$ of the equator. 
	Horizontal lines (large dots) mark the multi-model mean value for the control (forced) simulation.
	The smaller dots are representative of the values attained by individual models in the ensemble.}
	\label{fig:cmip_u}
\end{figure}

Putting together all the changes in eddy and mean meridional fluxes in the upper troposphere, all of the terms involved in the momentum budget decrease in magnitude, and the net effect on the annual and zonal mean zonal flow is a very slight decrease in the upper tropospheric easterlies (Figure \ref{fig:cmip_u}). It is interesting to note that the boreal summer zonal mean zonal flow is actually significantly weaker with warming, though still easterly in character. But, while the annual and zonal mean flow isn't affected to a large degree, there is a fair spread amongst the models that comprise the CMIP6 ensemble, and as Figure \ref{fig:cmip_u} suggests, the control has an easterly flow whereas some models in the CMIP6 ensemble indicate a switch with equatorial superrotation as an outcome of climate change.

\section{Conclusions and Discussion}

The zonal mean momentum budget of the upper tropics is known to primarily be a balance between two terms, eddy momentum flux convergence and mean meridional momentum flux convergence, that offset each other on seasonal and annual timescales \citep{lee1999climatological,dima2005tropical}.
Motivated by the longitudinal heterogeneity prevalent in the tropics, we have focused on the regional contributions that sum to make up the zonal mean.
Specifically, based on the zonal distribution of diabatic monsoonal heating, the nature of these fluxes is probed by splitting the tropics primarily into two regions. These are the Africa-Asia (Af-A; 30W - 150W) and the Central Pacific-West Atlantic (CP-WA; 150W - 30W) zones.

The Af-A sector provides the bulk of the eddy acceleration during both the solstitial seasons through the action of stationary Rossby waves present in that region throughout the year. The contribution from the CP-WA sector undergoes a semiannual reversal due to the disparate nature of the momentum fluxes prevalent here.
Akin to the Asian sector, divergent motions forced by the summer ITCZ, and the global monsoons, converge eddy fluxes of momentum into this region during summer. 
However, during winter, the CP-WA sector experiences a deceleration due to the cross-hemispheric transport of extra-tropical wave activity in the East Pacific.
In comparison, both the sectors contribute cohesively toward the zonally averaged mean flow deceleration term via the advection of absolute vorticity by divergent meridional winds. Hence, the two-way balance between zonal mean terms comprises of a seasonally sensitive and delicate three-way balance involving eddy fluxes of tropical and extratropical origin, and the mean meridional flux convergence.

Expanding width of the Tropics \citep{collins2013long, vallis2015response}, changing precipitation patterns \citep{ma2018responses, wang2020understanding} and weakening of the equatorial Pacific SST gradient \citep{dinezio2009climate, xie2010global} are some of the projections for a warmer Tropical climate, demonstrating that the impact of climate change is likely going to be felt strongly even at a regional level. With this in mind, the changes to the tropical momentum balance are analyzed in the second part of this study using data from the CMIP6 archive. Along with robust responses like the upward shift of the tropopause and decrease in the mean meridional flux due weakening Hadley Cell strength, we observe statistically significant changes in eddy fluxes in a warming scenario. These changes are mainly in the stationary component of the eddy flux, and in particular, we observe that eddy momentum fluxes are affected in Asian and East Pacific regions in summer and winter, respectively. 

In winter, over the East Pacific, the change in stationary waves causes a weakening of the upper tropospheric westerlies and almost shuts off cross-equatorial propagation of wave activity. This results in a decrease in magnitude of zonal deceleration experienced by this region. It is interesting to note that apart from influencing tropical momentum fluxes, this change in the westerly duct is likely to affect the intrusions of high potential vorticity air and tracers into the deep tropics \citep{waugh-polvani}. In contrast, during summer, changes are seen in both the upper troposphere and the lower stratosphere. In fact, flow around the Indian monsoon anticyclone weakens and strengthens in the upper troposphere and lower stratosphere, respectively. The former (latter) leads to a smaller (larger) eddy flux convergence over the equatorial Indian Ocean. In all, the lower stratospheric increase is larger, and thus there is a greater momentum flux convergence over the equatorial Indian Ocean with warming. Given that the Asian summer anticyclone plays an important role in the breaking of Rossby waves at the subtropical tropopause \citep{postel}, these structural changes could influence the frequency and intensity of such events as well as stratosphere-troposphere tracer and mass exchange \citep{chen1995,dunkerton1995}. Further, as pointed out by \cite{postel}, the Asian summer anticyclone appears as a smooth large-scale system in monthly and seasonal averages. In reality, it shows variability on finer spatial and temporal scales, and it would be interesting to investigate the influence of warming on this detailed structure.

The decrease in strength of upper tropospheric stationary waves is consistent with expectations from warming scenarios \citep[see, for example,][]{willis} and physical reasoning using an interpretation of zonal asymmetric waves as Rossby gyres \citep{LB1,LB2}. In addition, here, in the warming scenario, we observe a summer-time increase in strength of stationary waves in the lower stratosphere throughout the global tropics and subtropics. Indeed, during summer, rotational flow around streamlines in the eastern and western hemisphere subtropics and tropics increases in magnitude. This increase in magnitude of zonally asymmetric anomalies adds to projections of changes in stationary waves in the northern hemisphere upper stratosphere during boreal winter \citep{wang1} and in the southern hemisphere \citep{wang2}, and also coincides with a projected strengthening of the lower stratospheric Brewer-Dobson circulation \citep{butchart,butchart1}. This coincidence may be mediated by the "tropospheric control" pathway that links stationary wave amplitudes to the lower stratospheric zonal mean overturning circulation \citep{gerber}. Further, it is possible that the increased mass flux to the stratosphere in the subtropics during boreal summer \citep{deckert} comes from the Indian monsoon, and in turn this could mediate a stronger lower stratospheric Asian anticyclone by via localized divergence. In fact, the increased strength of divergence in the Sverdrup balance of the lower stratospheric Asian summer anticyclone is noticeable in the warming scenario.

Taken together, one aspect of this of this study showcases the extent of internal compensations that occur when upper equatorial momentum fluxes are studied in a zonal mean sense.
By presenting the results for the control simulations of CMIP6 alongside those for reanalysis, we hope to aid modelling groups in closing the gap between the two, and perhaps, provide better future projections on global and regional scales \citep{hall2014projecting, xie2015towards}. For example, the smaller magnitude of eddy acceleration in the summer in the control set, especially in the CP-WA region, as compared to present-day estimates possibly points to issues with the organization of tropical convergence zone in the model runs. The second set of results involve state-of-the-art projections that suggest significant changes to the dominant terms that contribute to the upper tropical momentum budget. At present, the balance of these terms is oriented to give upper tropical  zonal mean easterlies \citep{lee1999climatological, dima2005tropical}. Which way this balance tilts will affect the future state of the atmosphere, including potential tipping to superrotation as is thought to have prevailed in past climates \citep{pierrehumbert2000climate,tziperman2009pliocene}, both in a zonal mean sense and in terms of the regional climate. Indeed, projections suggest a decrease in magnitude of all terms involved in this budget and the ensemble mean zonal mean zonal flow in the upper equatorial troposphere in the warming runs is still marginally easterly in character, but with individual members showing both westward and superrotating flows.

%\section{References}

\bibliographystyle{apalike}
\bibliography{ref.bib}
	
\end{document}

\maketitle

\begin{figure}[ht!]
	\centering
	\includegraphics[width=0.9\linewidth]{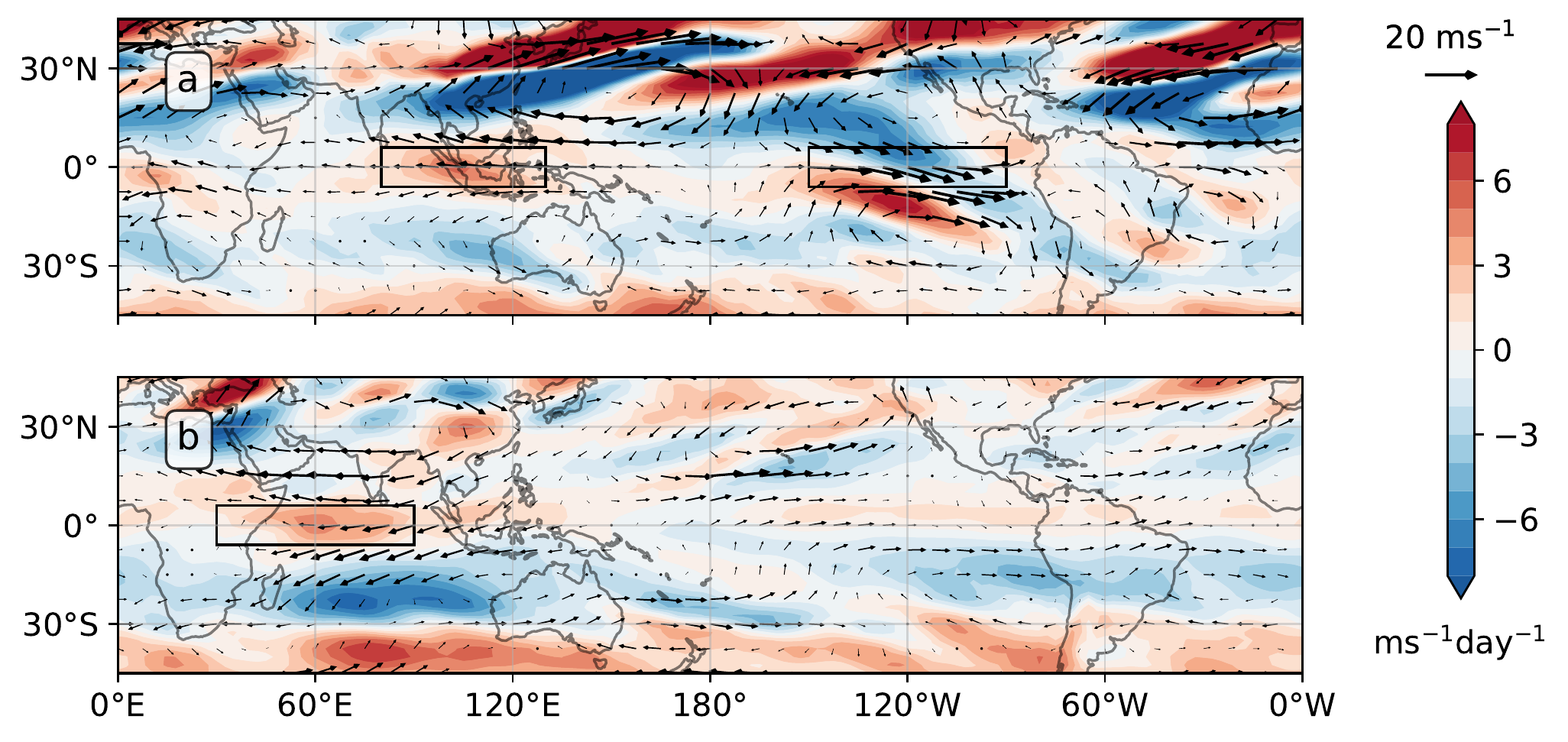}
	\caption{Same as Figure 4 in the text, except for the quivers which are for the eddy wind vectors. }
	\label{fig:eddiesClimWinSum}
\end{figure}

\begin{figure}[ht!]
	\centering
	\includegraphics[width=\linewidth]{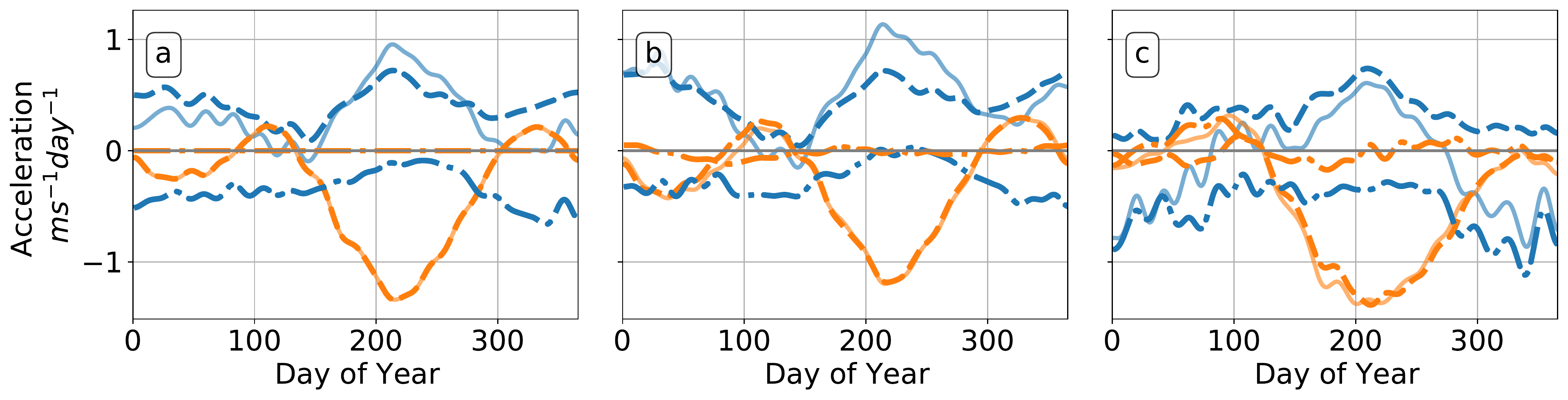}
	\caption{
		Same as Figures 1 and 3 except that dashed (dash-dot) lines represent the $u_r v_d$ ($u_r v_r$) eddy and mean momentum flux convergences (Equation 2) computed over a) all longitudes, b) Asia - Africa, c) Central Pacific - West Atlantic sectors. The lighter solid lines in each panel represent the full eddy and mean flux convergences over each of these regions.}
\end{figure}

\begin{figure}[ht!]
	\centering
	\includegraphics[width=0.7\linewidth]{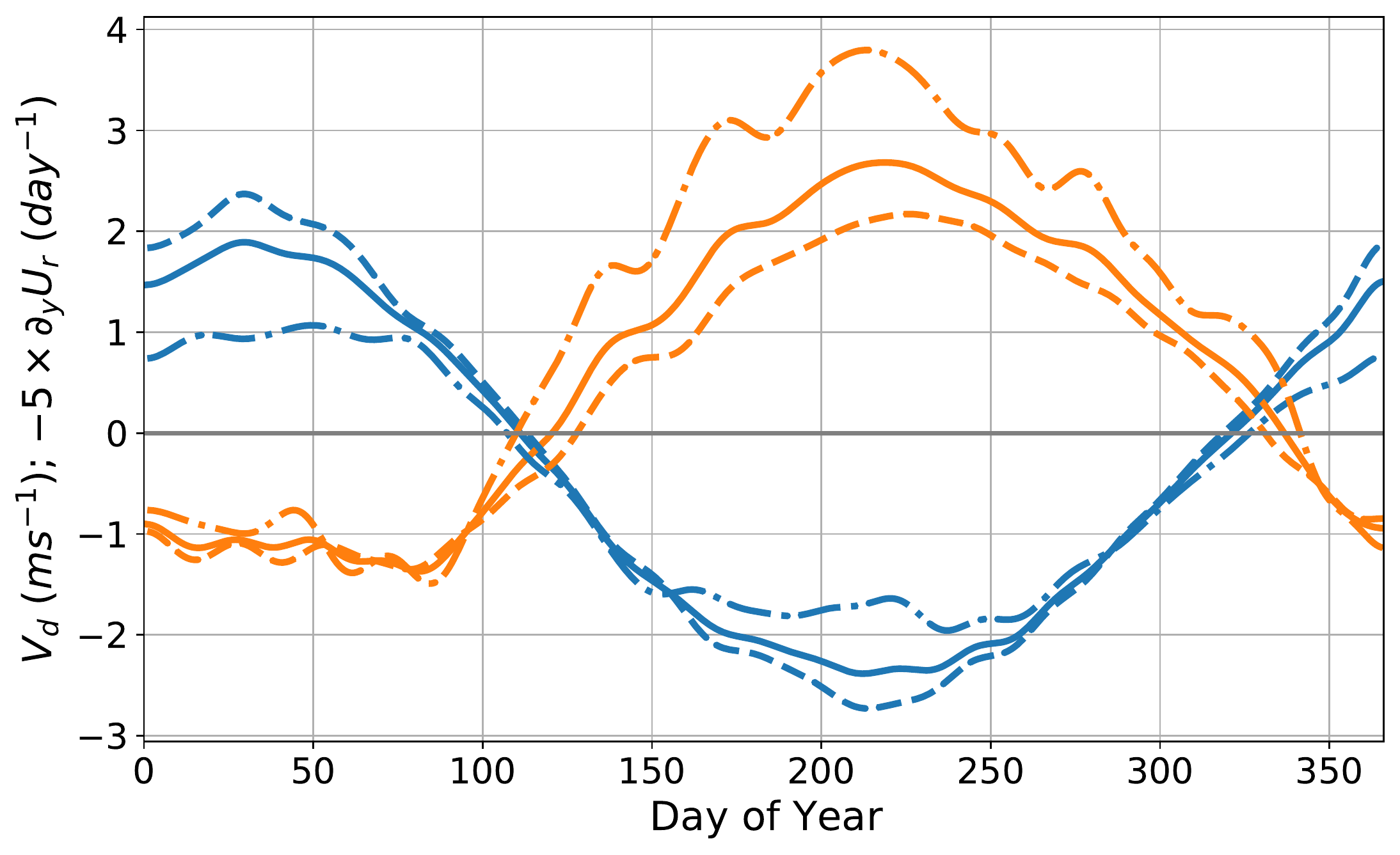}
	\caption{
		Climatological Day of Year variation of $v_d$ (blue) and $-5 \times \partial_y u_r$ (orange) zonally averaged over all longitudes (solid), A-Af (dashed), CP-WA (dash-dotted) and Africa (dotted). As Figures 1 and 3, all quantities are averaged over 150-300 hPa, about $\pm 5^{\circ}$. A 20-day low-pass filter has been applied prior to presentation. }
\end{figure}

\begin{figure}[ht!]
	\centering
	\includegraphics[width=\linewidth]{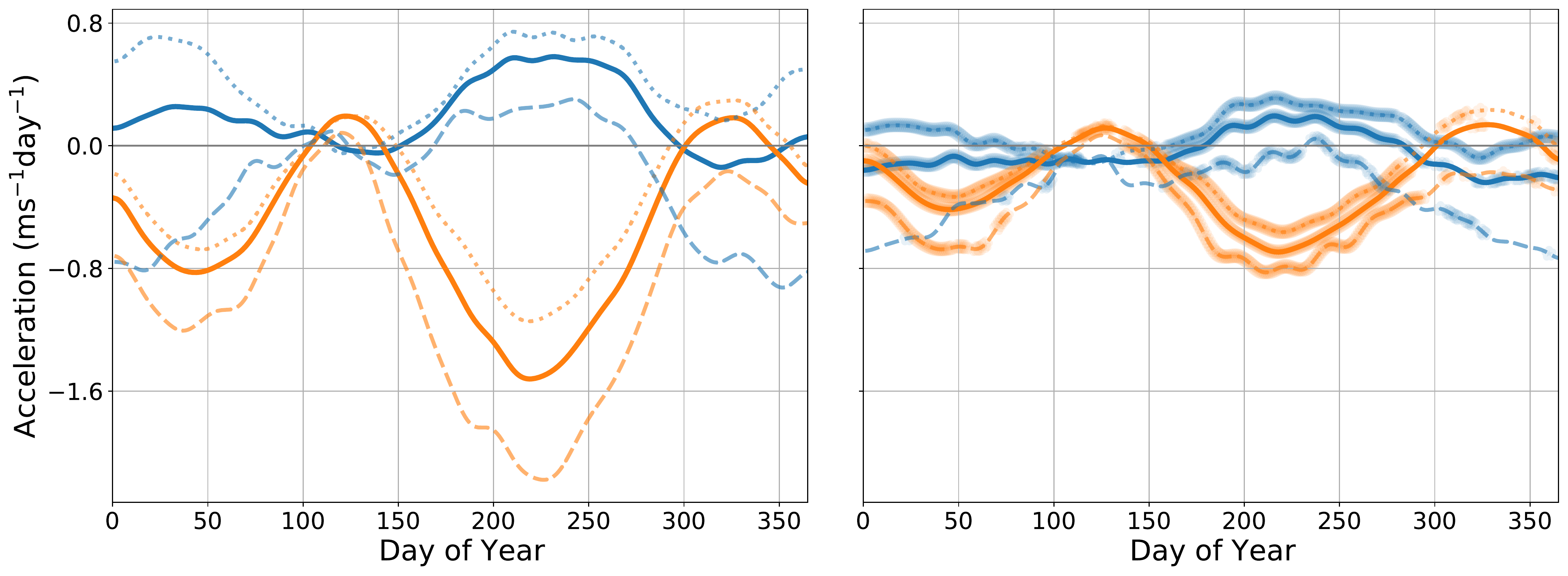}
	\caption{
		Same as Figure 3 in the text, except for the multi-model mean representations of the CMIP6 control (left) and forced simulations (right) at 250 hPa.
		Blurring of the curves, in the right hand panel, indicates days where the quantities related to the forced simulations are statistically different from those for control at the 95\% level of significance.}
	\label{fig:cmip2zones}
\end{figure}

\begin{figure}[ht!]
	\centering
	\includegraphics[width=0.9\linewidth]{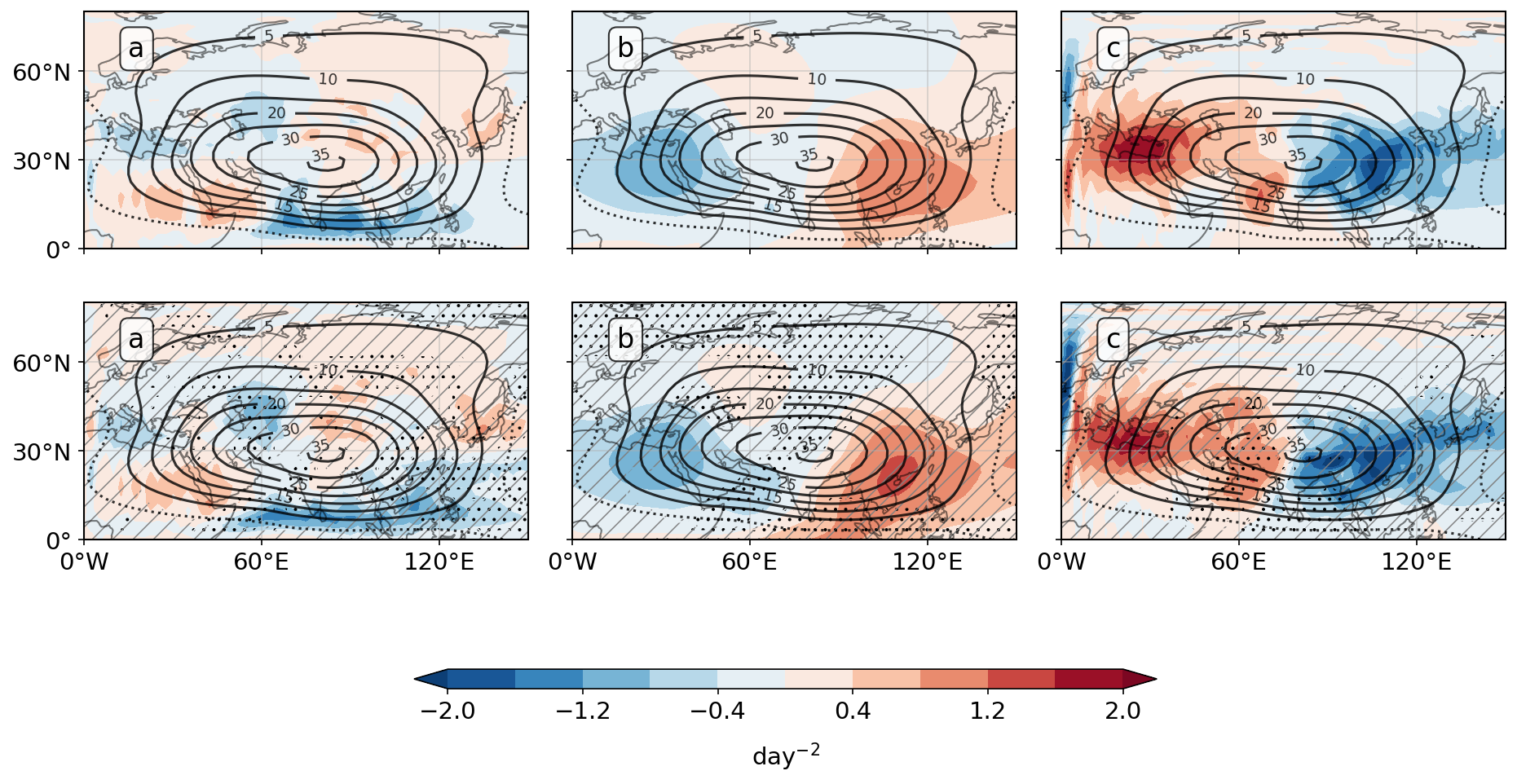}
	\caption{
		Boreal summer seasonal mean lower stratospheric (100 hPa) dominant terms of the vorticity budget %\citep[colours; see][]{sardeshmukh_held} and
		(colours) eddy streamfunction (contours) for control (top) and forced (bottom) simulations.
		Terms of the vorticity budget are a) $-\vec v \cdot \nabla \zeta$, b) $- \beta v$ and c) $ - (\zeta+f) \nabla \cdot \vec v$.
		Contour intervals are the same in all panels and the units are $10^6$ m$^2$s$^{-1}$.
		Hatching with gray lines (stippling with black dots) denotes changes in streamfunction (vorticity budget terms) that are statistically significant at the 95\% level by a two-tailed t-test.
	}
\end{figure}

\begin{figure}[ht!]
	\centering
	\includegraphics[width=0.8\linewidth]{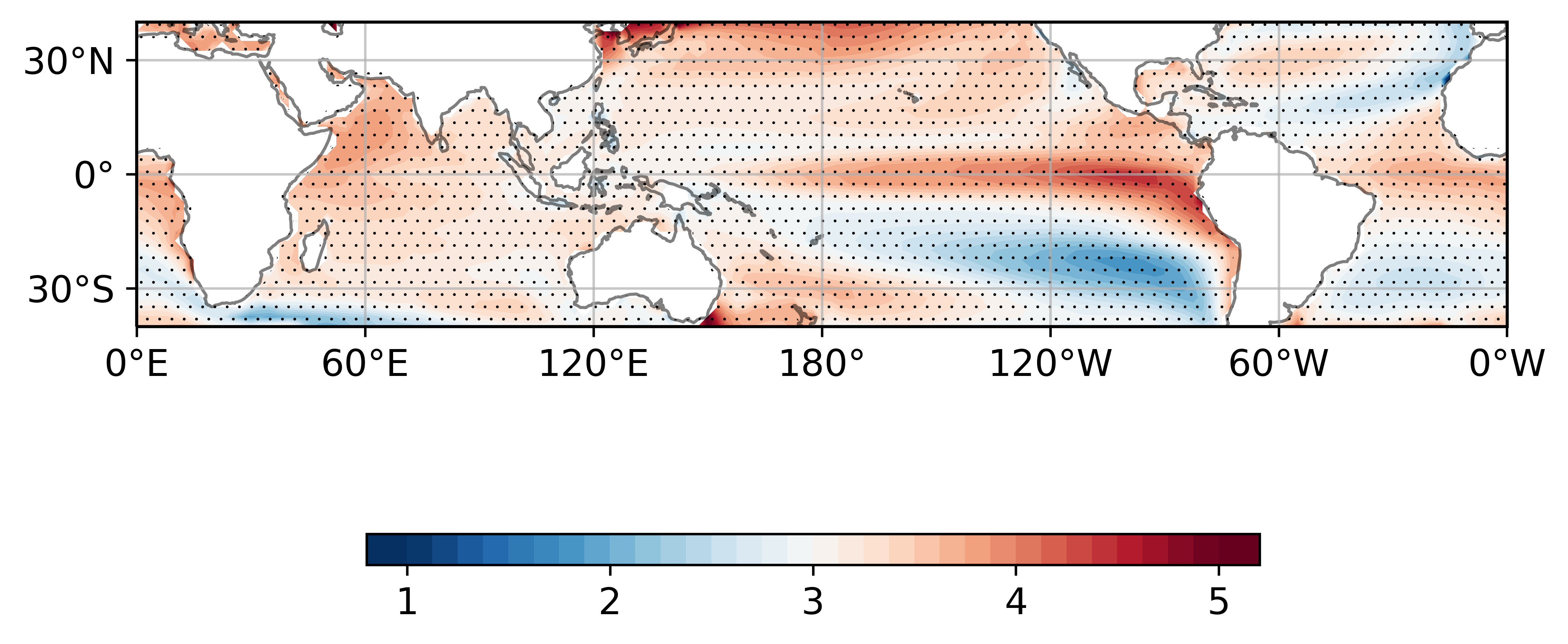}
	\caption{
		Time averaged multi-model mean difference in sea surface temperature (K) between forced and control simulations.
		Stippling denotes anomalies that are statistically significant at the 95\% level of significance.}
	\label{fig:cmip_sst_supp}
\end{figure}

\begin{figure}[ht!]
	\centering
	\includegraphics[width=\linewidth]{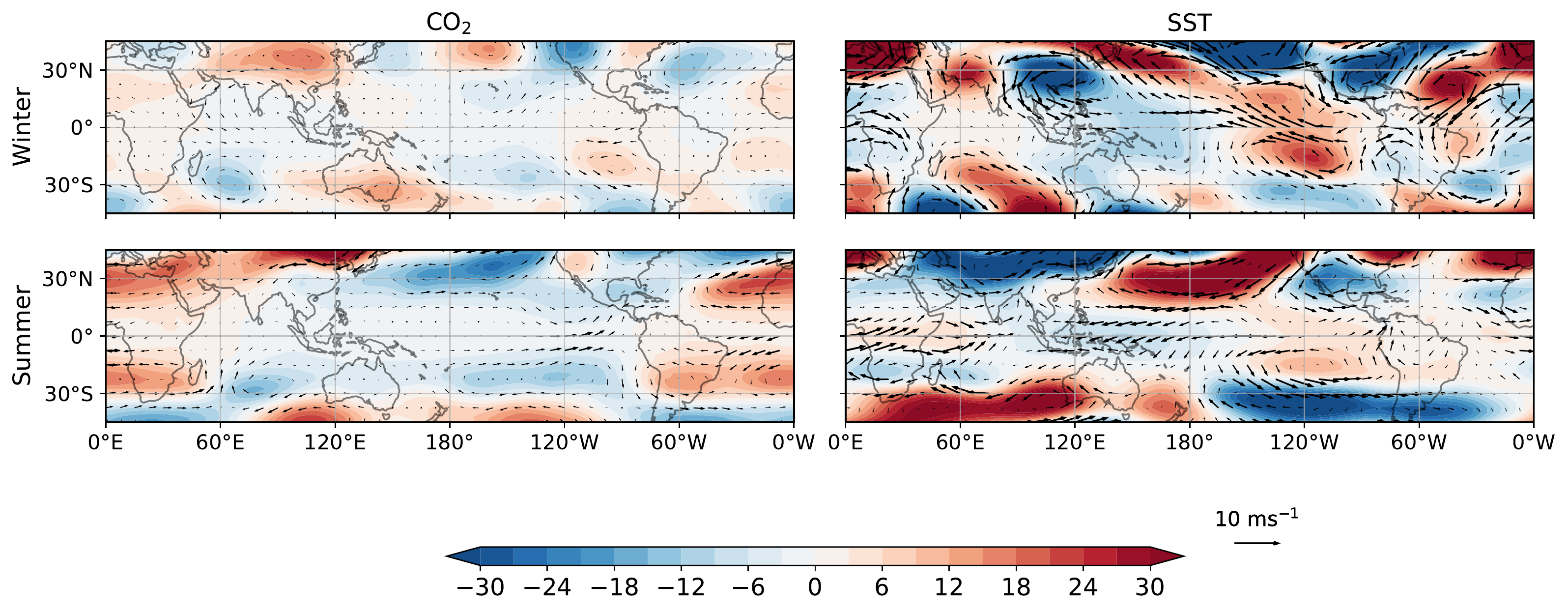}
	\caption{
		Seasonally averaged multi-model mean difference between forced and control simulations for eddy geopotential height (colors; m) and eddy horizontal wind velocity (quivers) at 250 hPa. 
		The data used here is from the CFMIP (Webb et al. 2017) available in the CMIP6 archive.
		We use data generated from 3 atmosphere-only scenarios by 5 different models.
		The scenarios are \textit{amip} (as control), \textit{amip-future4K} (as SST only) and \textit{amip-4xCO2} (as CO2 only) and the models are BCC-CSM2-MR, CESM2, MIROC6, MRI-ESM2-0 and IPSL-CM6A-LR .
		The results presented here are SST - Control (labelled as $SST$), and CO2 - Control (labelled as ${CO}_2$).
	}
	\label{fig:amip_zg_anom}
\end{figure}